\documentclass[amsmath,amssymb,pre,showpacs,floatfix,twocolumn]{revtex4}
\usepackage{graphicx}
\usepackage{dcolumn}

\begin{document}

\title{Rectifying the thermal Brownian motion of three-dimensional asymmetric objects}
\author{M. van den Broek}
\author{C. Van den Broeck}
\affiliation{Hasselt University, B-3590 Diepenbeek, Belgium}
\begin{abstract}
We extend the analysis of a thermal Brownian motor reported in
Phys.~Rev.~Lett.~\textbf{93}, 090601 (2004) by C. Van den Broeck, R. Kawai, and P. Meurs to a three-dimensional configuration. 
We calculate the friction coefficient, diffusion coefficient, and drift velocity as a function of shape and present estimates based on physically realistic parameter values.
\end{abstract}

\pacs{05.70.Ln, 05.40.Jc, 51.10.+y, 51.20.+d} 
\keywords{Brownian, motor, rectification, friction coefficient, diffusion coefficient, kinetic theory}

\maketitle

\section{Introduction}

Spectacular advances in bio- and nanotechnology make it possible, not only to measure or observe, but also to manipulate and construct objects at a very small scale. At the same time there is growing interest in techniques which can add functionality. In particular, the development of molecular engines is  a theme which has received great attention over the last two decades. The appearance of fluctuations in small systems has led to new concepts for characterizing or operating such devices, exploiting rather than fighting these very same fluctuations. 

These so-called Brownian motors \cite{Julicher,Reimann} have an additional theoretical interest through their relation with the old  issue of Maxwell demons  and the second law of thermodynamics. In return, this theoretical connection allows to make statements on the efficiency of such engines \cite{vandenbroeckefficiency}  or to transform them from engines into mini-refrigerators \cite{vandenbroeckrefrigerator,martijn}.
Most of the studies on Brownian motors start with an ad hoc separation of systematic and noise terms, based on linear Langevin equations.  This approach however offers little insight into the origin of the rectification of random fluctuations. 
As pointed out by van Kampen \cite{vankampen}, the rectification of nonlinear fluctuations cannot be addressed starting from the standard Langevin description with additive Gaussian white noise. 
In \cite{vandenbroeckmotor,meurs} a theoretical and numerical study of a thermal engine is presented in which rectification arises at the level of nonlinear response.
The analysis therein starts from a microscopic description based on Newton's laws of motion. There is another related distinct feature of the model: the asymmetry of the thermal engine  lies in the geometry of the motor itself, in contrast to the asymmetry imposed by the application of an external potential, appearing in the so-called flashing and rocking ratchet models. 

The characteristic properties of the engine, such as the friction coefficient, the speed and the diffusion coefficient, are calculated exactly in \cite{vandenbroeckmotor,meurs} and are found to be in excellent agreement with the results from hard disk molecular dynamics. 
However, the results are reported in dimensionless units, in part due to the fact that the analysis was, for reasons of simplicity and for comparison with molecular dynamics, limited to the case of two dimensions.  In view of the technological interest of motors in bio- and nanotechnology, we report here a full and detailed analysis of the three-dimensional version.

This paper is organized as follows. First, the model, notations, and working hypothesis are introduced in section \ref{model}.
The calculation method, based on the kinetic theory of gases, is presented in section \ref{kinetictheory}, with a discussion of the analytical solution following in section \ref{solution}. Finally, in section \ref{results} we report and discuss the results for the friction coefficient, diffusion coefficient, and drift velocity as a function of shape and present estimates based on physically realistic parameter values.

\section{The model}
\label{model}

The model presented in \cite{vandenbroeckmotor} reproduces in  a simplified way the principle ingredients of Feynman's ratchet and pawl mechanism \cite{feynman}: a temperature difference between two reservoirs and the presence in at least one reservoir of an asymmetric object. 
\begin{figure}
\includegraphics[width=0.7\columnwidth]{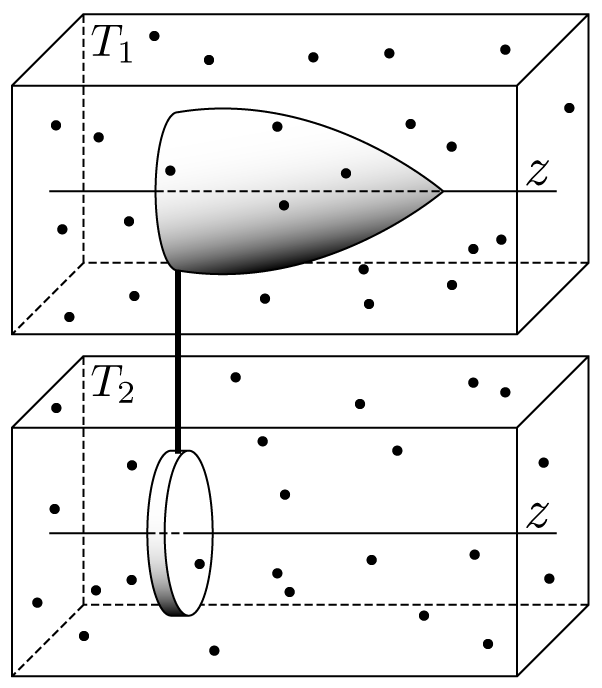}
\caption{The two-reservoir model of the thermal engine: two solid objects are confined in separate containers (scaled down for illustration purposes) that contain gases at temperatures $T_{1}$ and $T_{2}$. The objects are assembled with a rigid connection. The ensemble can move freely along the $z$-axis.}
\label{fig:model}
\end{figure}
The construction, extended to the case of three spatial dimensions, is as follows. We consider any number of reservoirs (denoted by index $i$), each containing a gas at equilibrium at  a temperature $T_{i}$. Fig.~\ref{fig:model} gives a schematic picture of the two-reservoir system. Solid objects with no internal degrees of freedom, called `motor units', are located inside the containers. These objects are coupled rigidly to each other, so that the motor moves as a single entity, with total mass $M$, along a given straight axis, corresponding to its single translational degree of freedom. For simplicity, we disregard any rotational degree of freedom (for a detailed discussion of this case, see \cite{martijn}).

Due to collisions with the gas particles (mass $m$), 
the motor will change its velocity $V(t)$
in the course of time $t$. The statistics of these collisions can be described under the assumption of molecular chaos, which is valid when the gases are in the high Knudsen number regime and the containers are large enough to avoid acoustic and other boundary effects. In addition, the shape of the units' surfaces must be such that no re-collisions with the motor occur, namely for convex and closed shapes. With these assumptions, the precollisional velocities are random and uncorrelated. Hence the time evolution of the probability $P(V, t)$ that the motor has speed $V$ at time $t$ can be described by a master equation:
\begin{equation}
\frac{\partial P(V, t)}{\partial t} = \int dV' \left[W(V|V') P(V', t) - W(V'|V) P(V, t)\right].
\label{master}
\end{equation}
Here $W(V|V')$ represents the transition probability per unit time for the motor to change its speed from $V'$ to $V$.

\section{Kinetic theory}
\label{kinetictheory}

In this section we study the collisions of  gas particles from either temperature reservoir with a motor part and derive the resulting total transition probability $W(V | V')$ for the motor to change speed from $V'$ to $V$.
We introduce a Cartesian coordinate system $(x, y, z)$ where the $z$-axis points along the free direction of movement of the motor.

\subsection{Conservation rules}
 A gas particle  will, upon collision with a motor unit, undergo an instantaneous change of velocity from $\vec{v'} = (v'_{x}, v'_{y}, v'_{z})$ before collision to $\vec{v} = (v_{x},v_{y}, v_{z})$ afterwards. 
Due to conservation of momentum along the free $z$-direction, one has:
\begin{equation}
m v'_{z} + M V' = m v_{z} + M V. \label{eq:conservationzmomentum}
\end{equation}
In addition, when the collision is perfectly elastic, the total energy is conserved:  
\begin{multline}
\frac{1}{2}M{V'}^2 + \frac{1}{2}m {v'_{x}}^2 + \frac{1}{2}m {v'_{y}}^2 +\frac{1}{2}m {v'_{z}}^2\\
= \frac{1}{2}MV^2 +\frac{1}{2}m v_{x}^2 + \frac{1}{2}m v_{y}^2 + \frac{1}{2}m v_{z}^2. \label{eq:conservationenergy}
\end{multline}
We will also suppose that the collision is described in terms of a (short-range) central force, implying that the
component of the momentum of the gas particle along any direction tangential to the surface of the motor is conserved.
\begin{figure}[h]
\includegraphics[width=0.5\columnwidth]{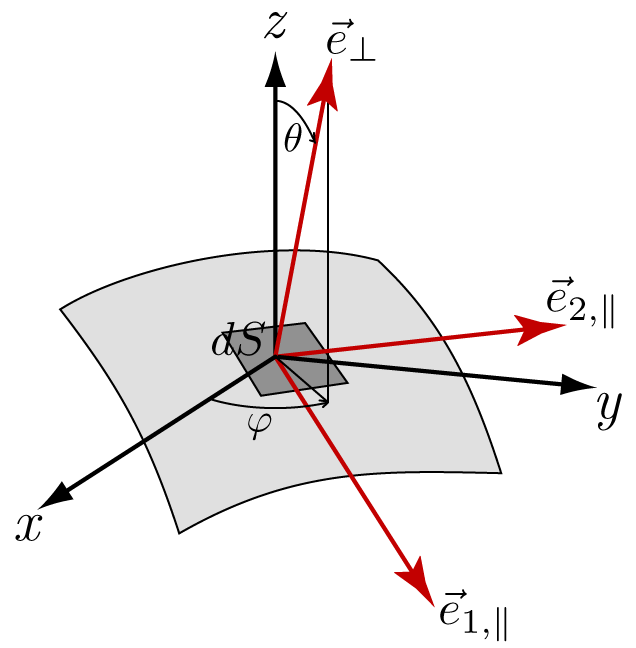}
\caption{(Color online) The orientation of an infinitesimal surface element $dS$ is represented by an outer-pointing unit normal vector $\vec{e}_{\perp}$ and determined by the spherical coordinates $\varphi$ and $\theta$. The polar angle $\theta$ is measured from the $z$-axis, which is chosen as the free direction of movement of the motor. $x, y$ complete the Cartesian coordinate system and the azimuthal angle $\varphi$ starts from the $x$-axis. Two orthogonal unit vectors $\vec{e}_{1,\shortparallel}$ and $\vec{e}_{2,\shortparallel}$ determine the plane that is tangent to the motor unit in  $dS$.}
\label{fig:coordinates}
\end{figure}
The orientation of the tangent plane to the motor surface is determined uniquely by a normal outward vector on an infinitesimal element $dS$ of the surface at the point of collision. In spherical coordinates, this normal vector is given by
\begin{equation}
\vec{e}_{\perp}\big{|}_{\text{cart}} \doteq (\sin \theta \cos\varphi, \sin \theta \sin \varphi,\cos \theta),
\end{equation}
with $\theta$ the polar angle from the $z$-axis ($0 \leqslant \theta \leqslant \pi$) and $\varphi$ the azimuthal angle in the $xy$-plane from the $x$-axis ($0 \leqslant \varphi < 2\pi$), cf. Fig.~\ref{fig:coordinates}.
We also introduce two mutually perpendicular unit vectors within the tangent plane:
\begin{align}
\vec{e}_{1,\shortparallel} &\doteq (\cos \theta \cos \varphi, \cos \theta\sin \varphi, -\sin \theta),\\
\vec{e}_{2,\shortparallel} &\doteq (-\sin \varphi, \cos \varphi, 0),
\end{align}
so that we can write the conservation of tangential momentum as
\begin{align}
\vec{v'} \cdot \vec{e}_{1,\shortparallel} & = \vec{v} \cdot
\vec{e}_{1,\shortparallel},\label{eq:conservationtang1}\\
\vec{v'} \cdot \vec{e}_{2,\shortparallel} & = \vec{v} \cdot
\vec{e}_{2,\shortparallel}.\label{eq:conservationtang2}
\end{align}
Solving the conservation rules [Eqs.~(\ref{eq:conservationzmomentum},\ref{eq:conservationenergy},\ref{eq:conservationtang1},\ref{eq:conservationtang2})]
for $V, v_{x},v_{y}, v_{z}$ leads to the following expression for the postcollisional speed $V$ of the motor in terms of the precollisional speeds:
\begin{align}
V &= V'  \\ 
&+ \frac{2 \frac{m}{M} \cos^2 \theta}{1 + \frac{m}{M} \cos^2 \theta} (v'_{x} \tan \theta \cos \varphi + v'_{y}\tan \theta \sin\varphi + v'_{z} - V'). \label{eq:postcollisional}
\end{align}

\subsection{Transition probability}
The motor is  subject to random collisions by gas particles form the different reservoirs $i$. The particle density and velocity distribution in reservoir $i$ are denoted by $\rho_{i}$ and $\phi_{i}(v_{x}, v_{y}, v_{z})$, respectively. The contribution $dW_{i}(V | V')$ to the total transition probability $W(V | V')$, coming from collisions on an infinitesimal surface element $dS_{i}$ of the motor unit in reservoir $i$,  can then be found by considering the number of gas particles that collide with this surface element $dS_{i}$ in a unit time step:
\begin{align}
dW_{i}&(V | V') = dS_{i} \int_{-\infty}^{+\infty} dv'_{x} \int_{-\infty}^{+\infty} dv'_{y} \int_{-\infty}^{+\infty} dv'_{z} \nonumber \\
&\times  H[(\vec{V'} - \vec{v'}) \cdot \vec{e}_{\perp}] |(\vec{V'} - \vec{v'}) \cdot \vec{e}_{\perp}| \rho_{i} \phi_{i}(v'_{x}, v'_{y}, v'_{z}) \nonumber\\
&\times \delta \big{[}V - V' \nonumber\\
& - B(\theta) (\tan \theta \cos \varphi \medspace v'_{x} + \tan \theta \sin
\varphi \medspace v'_{y} + v'_{z} - V')
\big{]}.
\label{eq:dW}
\end{align} 
Here $H$ represents the Heaviside function, $\delta$ the Dirac distribution, and
\begin{equation}
B(\theta) = \frac{2 \frac{m}{M} \cos^2 \theta}{1 + \frac{m}{M} \cos^2 \theta}.
\end{equation}
The Dirac delta distribution selects those particles that
produce the required postcollisional speed $V$
, following the collision rules [Eq.~(\ref{eq:postcollisional})]. 
Assuming that the velocity distribution of the gas particles is Max\-well\-ian at the reservoir temperature $T_{i}$,
\begin{equation}
\phi_{i}(v_{x}, v_{y}, v_{z}) = 
\left(\frac{m}{2 \pi k_{B} T_{i}}\right)^{3/2}
\exp \left[- \frac {m (v_{x}^2 + v_{y}^2 + v_{z}^2)}{2 k_{B} T_{i}}\right],
\end{equation}
the integrals over $v'_{x}, v'_{y}, v'_{z}$ in Eq.~(\ref{eq:dW}) can be calculated explicitly. The total transition probability for the motor to change velocity from $V'$ to $V$ in a unit time, is then found by integrating $dW_{i}(V | V')$ over the surface $S_{i}$ of each motor part and summing over all the reservoirs $i$:
\begin{multline} W(V | V') = 
\frac{1}{4} \sum_{i}
 \rho_{i}
\sqrt{\frac{m}{2 \pi k_{B} T_{i}}}\\
\times \Bigg{(} 
  (V - V') H[V - V'] \int_{S_{i},\cos\theta>0} d S_{i} 
\\
+ (V' - V) H[V' - V] \int_{S_{i},\cos\theta<0} d S_{i}
\Bigg{)}
\left(\frac{M}{m \cos \theta} + \cos \theta \right )^{2}
\\
\times \exp \left[- \frac{m}{2 k_{B} T_{i}}
\cos^{2} \theta
\left(V' + \frac{1}{2} (1 + \frac{M}{m \cos^{2} \theta}) (V - V') \right )^{2}
\right ]. \label{transitionprobability}
\end{multline}

We note that in the case of a single reservoir or for multiple reservoirs at the same temperature, the transition probability satisfies the following relation:
\begin{equation}
W(V | V') P^{eq}(V')=W(-V' | -V) P^{eq}(-V),
\end{equation}
with $P^{eq}(V)$ the Maxwell Boltzmann distribution for the speed $V$ of the motor unit (at the temperature of the reservoir(s)). This is in agreement with the general principle of detailed balance in a system at equilibrium \cite{onsager}.

\section{Solution method}
\label{solution}

We follow the method of \cite{meurs} to solve the master equation [Eq.~(\ref{master})] for the moments of the motor velocity, 
\begin{equation}
\langle V^{n} \rangle = \int_{-\infty}^{\infty} P(V, t) V^{n} dV.
\end{equation}
This method is based on the van Kampen $1 / \Omega$ expansion \cite{vankampen}.

\subsection{Solution of the master equation}
It is convenient to scale the motor velocity $V$ to a dimensionless variable
\begin{equation}
X = \sqrt{\frac {M}{k_{B} T_{\text{eff}}}} V,
\end{equation}
with the effective temperature $T_{\text{eff}}$ to be determined self-consistently from the condition $\langle X^{2} \rangle = 1$ in steady state operation.
We can expand the integrand in Eq.~(\ref{master}) in a Taylor series about $X'$:
\begin{equation}
\frac{\partial P(X, t)}{\partial t}  = \sum_{n = 1}^{\infty} \frac{(-1)^{n}}{n!}
\frac{d^{n}}{dX^{n}}\{J_{n}(X) P(X, t) \}.
\label{derivP}
\end{equation}
Here the `jump moments' are given by
\begin{equation}
J_{n}(X) = \int {\Delta X}^{n} W(X; \Delta X) d\Delta X,
\label{jumpmoments}
\end{equation}
with $W(X'; \Delta X) = W(X | X')$ and $\Delta X = X' - X$.
Using Eq.~(\ref{derivP}) a coupled set of equations for the time evolution of the moments $\langle X^{n} \rangle$ can then be constructed:
\begin{equation}
\frac{\partial \langle X^{n} \rangle}{\partial t}  = \sum_{k=1}^{n} \binom {n}{k} \langle X^{n -
k} J_{k}(X) \rangle, \label{eq:moments}
\end{equation}
with $\binom {n}{k}$ the binomial coefficients.

The exact expression for the jump moments $J_{n}(X)$ is obtained by integration over $\Delta X$ in Eq.~(\ref{jumpmoments}).
In terms of parabolic cylinder functions \cite{gradshteyn},
\begin{multline}
\operatorname{D}_{n}(z) = \left( \exp[-z^{2}/4] / \Gamma[-n] \right) \\
\times \int_{0}^{\infty} \exp[-zx-x^{2}/2] x^{-n-1} dx \quad \text{(for $n < 0$),}
\end{multline}
and the Gamma function ($\Gamma$), the result reads: 
\begin{align}
J_{n}(X) = &
\frac{2^{n}}{\sqrt{2 \pi}}
\Gamma[n+2]
\left(\sqrt{\frac{M}{m}} \right)^{n}
\\
&\times \sum_{i} \rho_{i} \sqrt{\frac{k_{B} T_{i}}{m}}
\left(\sqrt{\frac{T_{i}}{T_{\text{eff}}}} \right)^{n}
\\
&\times \int_{S_{i}} dS_{i}
\left( \cos \theta + \frac{M}{m} \frac{1}{\cos \theta}\right)^{-n}
\\
&\times \exp\left[- \frac{1}{4} \frac{m}{M}
\frac{T_{\text{eff}}}{T_{i}} X^{2} \cos^{2} \theta \right ]
\\
&\times \operatorname{D}_{-n-2} \left[\frac{1}{2} \sqrt{\frac{m}{M}}
\sqrt{\frac{T_{\text{eff}}}{T_{i}}} X \cos \theta \right]. 
\end{align} 

\begin{table*}
\caption{\label{tabel:sigma}The geometric moments $\sigma_{2}$ and $\sigma_{3}$ for some basic three-dimensional shapes. The lowest orders $\sigma_{2}$ and $\sigma_{3}$ ($\sigma_{1} = 0$) are tabulated in terms of the total surface area $S$ of the geometry. For illustrations of the shapes and their parameters, see Fig.~\ref{fig:geometries}.}    
\begin{ruledtabular}
\begin{tabular}{llll}
Shape & $\sigma_{2} / S $ & $\sigma_{3} / S $ & Surface $S$ 
\\
\hline
Disk & $1$ & $0$ & $2 \pi r^{2}$
\\
Blade & $1$ & $0$ & $2 l w$
\\
Sphere & $1/3$ & $0$ & $4 \pi r^{2}$
\\
Cone & $\sin \alpha$ & $\sin \alpha (\sin \alpha - 1)$ & $\pi r^{2} (1 + \csc \alpha )$
\\
Pyramid & $\sin \alpha$ & $\sin \alpha (\sin \alpha - 1)$ & $\frac{1}{4} n r^{2} \cot \frac{\pi}{n} \left(1 + \csc \alpha \right)$
\\
Spherical cap & $\frac{\cos 2 \alpha + 5 \cos \alpha +6}{3 \cos \alpha + 9}$ & $\frac{\sin^{4} \alpha}{\cos 2 \alpha + 4 \cos \alpha - 5}$ & $\pi r^{2} (3 + \cos \alpha) / (1 + \cos \alpha)$
\\
Spherical cone & $\frac{3 \sin^{3} \alpha - 2 \cos^{3} \alpha + 2}{3 \sin \alpha - 6 \cos \alpha + 6}$ & $\frac{2 \sin^{4} \alpha + cos^{4} \alpha - 1}{4 \cos \alpha - 2 \sin \alpha - 4}$ & $\pi r^{2} \csc \alpha (1 + 2 \tan(\alpha/2))$
\end{tabular}
\end{ruledtabular}
\end{table*}
\begin{figure*}
\includegraphics[width=0.75\textwidth]{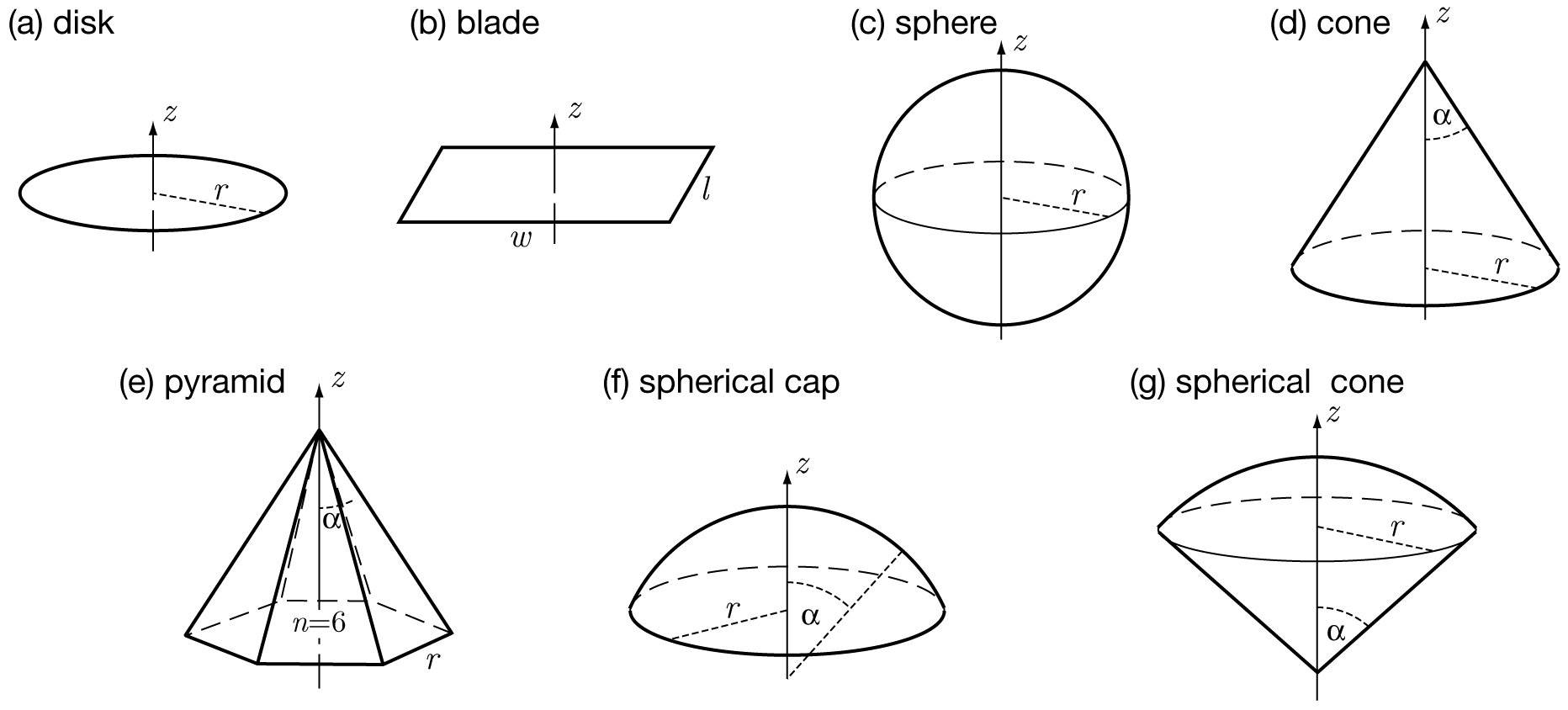}
\caption{Examples of simple three-dimensional geometries that can be treated analytically. The parameters that determine the relative areas of the surfaces are indicated in the illustrations.}
\label{fig:geometries}
\end{figure*}

An exact solution of Eq.~(\ref{eq:moments}) is not available.  We therefore turn to a perturbational approach  in terms of the parameter 
\begin{equation}
\varepsilon = \sqrt{m/M}.
\end{equation}
This is consistent with the observation that the mass $M$ of the motor is expected to be much larger than the mass $m$ of the gas particles. Even for a motor with dimensions of nanometers operating in a gaseous environment,  $\varepsilon$ is of order of $10^{-3}$.

The expansion of the parabolic cylinder functions is given by
\begin{multline}
2^{-n/2 }\Gamma[n+2] \operatorname{D}_{-n-2}(z)
= \Gamma[\frac{n+2}{2}]
- \sqrt{2} \Gamma[\frac{n+3}{2}] z \\
+ \frac{2 n + 3}{4} \Gamma[\frac{n + 2}{3}] z^2
- \frac{2n + 3}{6 \sqrt{2}} \Gamma[\frac{n +3}{2}] z^3 \\
+ \frac{4 n^2 + 12 n + 11}{96} \Gamma[(2 + n)/2] z^4 + O(z^{5}).
\end{multline}
Introducing a scaled time $\tau = \varepsilon^{2} t$, we find the following expansion for the equation of the first moment:
\begin{multline} 
\frac{\partial \langle X  \rangle}{\partial \tau}  =
\sum_{i} \rho_{i} \sqrt{\frac{k_{B} T_{i}}{m}}
\biggl[
\varepsilon^{-1} \sqrt{\frac{T_{i}}{T_{\text{eff}}}}  \sigma_{1,i}
- 2 \sqrt{\frac{2}{\pi}} \langle X  \rangle  \sigma_{2,i}
\\
+ \varepsilon \left(
\sqrt{\frac{T_{\text{eff}}}{T_{i}}} \langle X ^{2} \rangle
-\sqrt{\frac{T_{i}}{T_{\text{eff}}}}\right) \sigma_{3,i}
\\
+ \frac{ \varepsilon^{2}}{3} \sqrt{\frac{2}{\pi}} \left(6 \langle X  \rangle -
\frac{T_{\text{eff}}}{T_{i}}\langle X ^{3}
\rangle \right)  \sigma_{4,i}
 \\
+ \varepsilon^{3} \left(\sqrt{\frac{T_{i}}{T_{\text{eff}}}} - 
\sqrt{\frac{T_{\text{eff}}}{T_{i}}}
\langle X ^{2}
\rangle \right)   \sigma_{5,i}
\biggr]
+ O(\varepsilon^{4}). \label{firstmoment}
\end{multline}
The geometry of the motor is  contained in the shape factors $\sigma_{n,i}$, defined as:
\begin{equation}
\sigma_{n,i} = \int_{S_{i}} dS_{i} \cos^{n} \theta.
\label{eq:sigma}
\end{equation}
At this point we remark that the azimuthal angle $\varphi$ has dropped out, and that the geometric dependency is determined only by the (polar) angle $\theta$ between the surface and the direction of movement.
Also note that the term in $\varepsilon^{-1}$ in Eq.~(\ref{firstmoment}) is zero by application of Gauss' theorem:
\begin{equation}
\sigma_{1,i} = \int_{S_{i}} dS_{i} \cos \theta
= \int_{S_{i}} dS_{i} \vec{e}_{\perp} \cdot \vec{e}_{z}
= \int_{V_{i}} dV_{i} (\nabla \cdot \vec{e}_{z}) = 0,
\end{equation}
where the latter integral is over the interior volume $V_{i}$ of a motor part. This is consistent with the fact that there is no net macroscopic force acting on the motor. The net motion that will be revealed below is the effect of fluctuations only.

Similarly, for the equation of the second moment, one finds the following expansion:
\begin{multline} 
\frac{\partial \langle X^{2}  \rangle}{\partial \tau} =\sum_i  \rho_i\sqrt{\frac{k_B T_i}{m}}
 \biggl[-4\sqrt{\frac{2}{\pi}}
\left(-\frac{T_i}{T_{\text{eff}}}+\langle X^2\rangle\right)\sigma_{i}^2
\\ 
-2\varepsilon\left(4\sqrt{\frac{T_i}{T_{\text{eff}}}}\langle X\rangle-\sqrt{\frac{T_{\text{eff}}}{T_i}}\langle X^3\rangle\right)
\sigma_{i}^3 
\\ 
 +2\varepsilon^2\sqrt{\frac{2}{\pi}}\left(-4\frac{T_i}{T_{\text{eff}}}+5\langle X^2\rangle-\frac{1}{3}\frac{T_{\text{eff}}}{T_i}\langle X^4\rangle
\right)\sigma_{i}^4 
\biggr]
+O(\varepsilon^{3}).
\label{secondmoment}
\end{multline}

\subsection{Linear relaxation}
To order $\varepsilon^{0}$, Eq.~(\ref{firstmoment}) reduces to a linear relaxation law $M \partial_{t} \langle V \rangle = - \gamma \langle V \rangle$ with $\gamma = \sum_{i} \gamma_{i}$ the
sum of linear friction coefficients $\gamma_{i}$ of each part of the object:
\begin{equation}
\gamma_{i} = 4  \rho_{i} \sqrt{\frac{k_{B} T_{i} m}{2 \pi}}
\sigma_{2,i}
= \varrho_{i} \bar{v}_{i} \sigma_{2,i},
\label{eq:frictioncoefficients}
\end{equation}
where $\varrho_{i} = m \rho_{i}$ is the mass density of the gas, $\bar{v}_{i} = \sqrt{8 k_{B} T_{i} / (\pi m)}$ is the mean gas velocity, and $\sigma_{2,i}$ is a geometric factor.

\subsection{Nonlinearity: Steady state directed motion}
If a constant temperature difference between the reservoirs can be maintained for a time longer than the relaxation time $M/\gamma$, the probability distribution will relax to its steady state value. Restricting ourselves to the first two moments, we turn to the steady state solution of Eqs.~(\ref{firstmoment}) and (\ref{secondmoment}). First, from Eq.~(\ref{secondmoment}) we determine the effective temperature $T_{\text{eff}}$, which was defined earlier by the condition $\langle X^2\rangle = 1$. To lowest order, $\varepsilon^{0}$, we find that $T_{\text{eff}}$ is the weighted average of the reservoir temperatures:
\begin{equation}
T_{\text{eff}} = \frac{\sum_{i} \gamma_{i} T_{i}}{\sum_{i} \gamma_{i}}
=  \frac{\sum_{i} \rho_{i} \sigma_{2,i} T_{i}^{3/2}}{\sum_{i} \rho_{i} \sigma_{2,i} T_{i}^{1/2}}.
\end{equation}

Solving Eq.~(\ref{firstmoment}) to order $\varepsilon^{0}$ leads to a zero average drift speed $\langle X\rangle = 0$. Rectification of the thermal fluctuations occurs at higher levels of the expansion and it is necessary to include nonlinear terms. Solving Eq.~(\ref{firstmoment}) up to order $\varepsilon$ gives us an expression to lowest order for the average drift speed of the motor:
\begin{align}
\langle V \rangle &= \sqrt{\frac{m}{M}}
\sqrt{\frac{\pi k_{B} T_{\text{eff}}}{8 M}}
\frac{\sum_{i} \rho_{i} (\frac{T_{i}}{T_{\text{eff}}} - 1) \sigma_{3,i}}
{\sum_{i} \rho_{i} \sqrt{\frac{T_{i}}{T_{\text{eff}}}} \sigma_{2,i}}
\nonumber\\
&= \sqrt{\frac{m}{M}}
\sqrt{\frac{\pi k_{B}}{8 M}}
\frac
{ \sum_{i} \sum_{j} \rho_{i} \rho_{j} \sigma_{3,i} \sigma_{2,j} (T_{i} - T_{j}) \sqrt{T_{j}} }
{ (\sum_{i} \rho_{i} \sqrt{T_{i}} \sigma_{2,i})^{2} }.
\label{driftspeed}
\end{align}
 
\section{Results and discussion}
\label{results}
\subsection{Friction and diffusion coefficients}
\begin{figure}
\includegraphics[width=0.8\columnwidth]{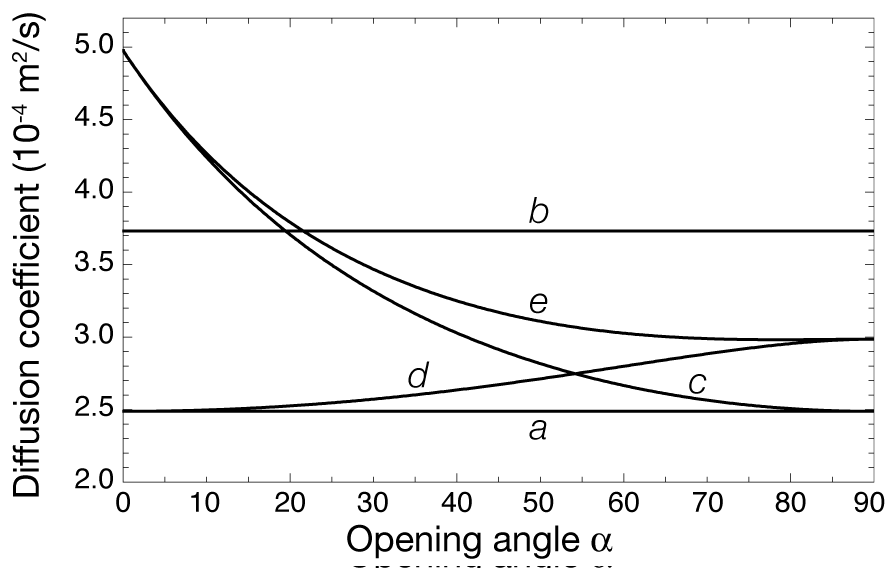}
\caption{Diffusion coefficients of objects with different geometry but identical cross section $\pi r^{2}$ ($r = 100$ nm) in highly diluted argon gas ($10^{19}\, \text{m}^{-3}$ particles, mean free path 2.3 $\mu$m) at 299 K temperature.
The geometries are
(a) a disk,
(b) a sphere,
(c) a cone,
(d) a spherical cap, and
(e) a spherical cone. See Fig.~\ref{fig:geometries} for illustrations. For (c,d,e) the shapes and hence the diffusion coefficients depend on an opening angle $\alpha$.}
\label{fig:diffusioncoefficients}
\end{figure}
Although not directly related to the main topic of this work, we briefly pause to discuss the new result for the linear friction coefficient, given in Eq. (\ref{eq:frictioncoefficients}). Together with results from Table \ref{tabel:sigma} for the geometric factors $\sigma_{2}$  this result provides the explicit expression for the linear friction coefficient of corresponding basic shapes. The result for a spherical shape with radius $r$,
\begin{equation}
\gamma
= \frac{4}{3} \pi r^{2} \varrho_{i} \bar{v}_{i},
\end{equation}
is in agreement with the result found in \cite{epstein}.
In combination with the Einstein relation $D = k_{B} T / \gamma$, we also obtain the explicit formulas for the corresponding diffusion coefficients $D$. As an illustration, numerical values are given in   Fig.~\ref{fig:diffusioncoefficients} for various shapes
of cross section $\pi r^{2}$, $r = 100$ nm. To be concrete, we consider highly diluted argon  gas (density $10^{19}\, \text{m}^{-3}$) leading to diffusion coefficients of the order of $3\times 10^{-4}\, \text{m}^{2} / \text{s}$, and corresponding friction coefficients of order $1.5\times 10^{-17}\, \text{Ns} / \text{m}$. As expected, the conical shapes have higher diffusion coefficients and lower friction as  one considers  smaller opening angles $\alpha$.

\subsection{Equilibrium}
When the reservoirs are all at the same temperature, $T_{i} = T$, we find $T_{\text{eff}} = T$, and the distribution of the moments is Gaussian:
\begin{equation}
\langle X \rangle = 0, \langle X^{2} \rangle = 1, \langle X^{3} \rangle = 0,\langle X^{4} \rangle = 3, \ldots
\end{equation}
The notion that it is impossible to achieve directed motion (or equivalently, to extract work) from a system in thermal equilibrium is confirmed. At least two  reservoirs at different temperature are necessary to break detailed balance and make possible the rectification of thermal fluctuations. 

\subsection{Asymmetry}
The geometry of the motor units enters into the expression of the average speed via de shape factors  $\sigma_{2}$ and $\sigma_{3}$. In table \ref{tabel:sigma}, we have reproduced these quantities for the objects depicted in Fig.~\ref{fig:geometries}. 
As is clear from symmetry arguments, the appearance of systematic motion in one direction requires, apart from non-equilibrium conditions, also the breaking  of the spatial symmetry in the system.
One easily verifies from Eq.~(\ref{eq:sigma}) that
\begin{equation}
\sigma_{n,i} = 0, \quad n \text{ odd},
\end{equation}
when the surface of a motor element possesses reflection symmetry along the $z$-axis, the direction of motion.
Consistent with this symmetry observation we find that the drift speed of the motor in steady state is indeed zero at lowest order in the perturbation when  $\sigma_{3,i} = 0$, cf.  Eq.~(\ref{driftspeed}). 
It is however interesting to note that reflection symmetry is a sufficient but not a necessary condition for $\sigma_{3,i} = 0$, and hence for obtaining zero sustained motion (at least in this order of the approximation). 
\begin{figure}
\includegraphics[width=0.8\columnwidth]{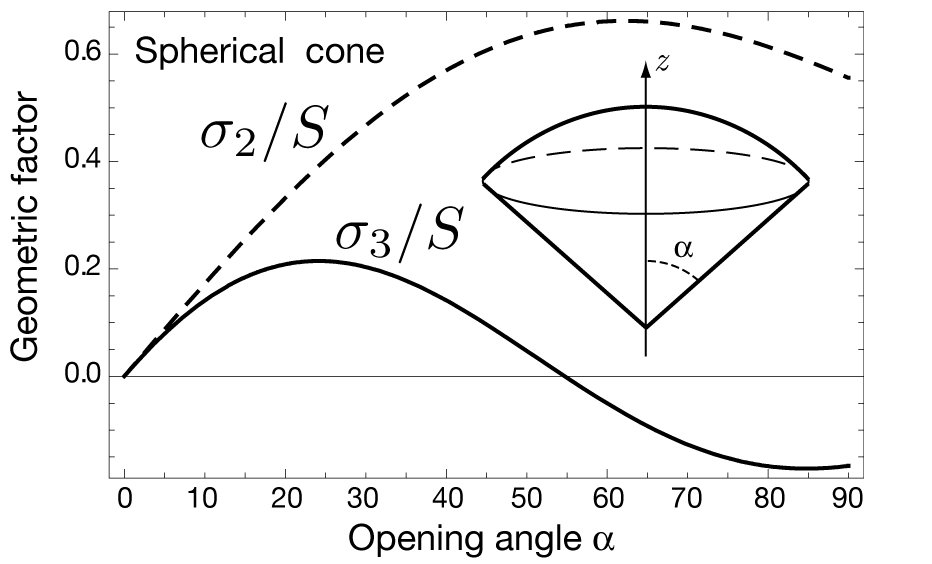}
\caption{The geometric factors $\sigma_{2} / S $ and $\sigma_{3} / S $ as a function of the opening angle $\alpha$ for a spherical cone, with $S$ the surface area. While $\sigma_{2}$ appears in the expression for the friction coefficient, $\sigma_{3}$ relates to the drift speed of the motor. When the geometry of the surface exhibits symmetry along the $z$-axis (the direction of movement), $\sigma_{3}$ is zero, and the motor shows no directed motion. The spherical cone is an example of a class of shapes for which $\sigma_{3}$ can become zero, and hence the drift speed (at least to the first approximation), without however showing reflection symmetry. For this particular case, $\sigma_{3} = 0$ for $\alpha \approx 55^{\circ}$.}
\label{fig:sphericalcone}
\end{figure}
Consider for example a spherical cone, see Fig.~\ref{fig:sphericalcone}. For a specific opening angle of $\approx 55^{\circ}$ $\sigma_{3,i}$ becomes zero, even though there is no reflection symmetry.
A similar  discussion can be applied to higher orders corrections in the $\varepsilon$-expansion (featuring the appearance of the higher shape factors $\sigma_{5}$, $\sigma_{7}$, and so on) indicating that there are special shapes which will have a very low average speed even though there are no immediate symmetry reasons to expect so.

\subsection{Temperature gradient}
\begin{figure}
\includegraphics[width=\columnwidth]{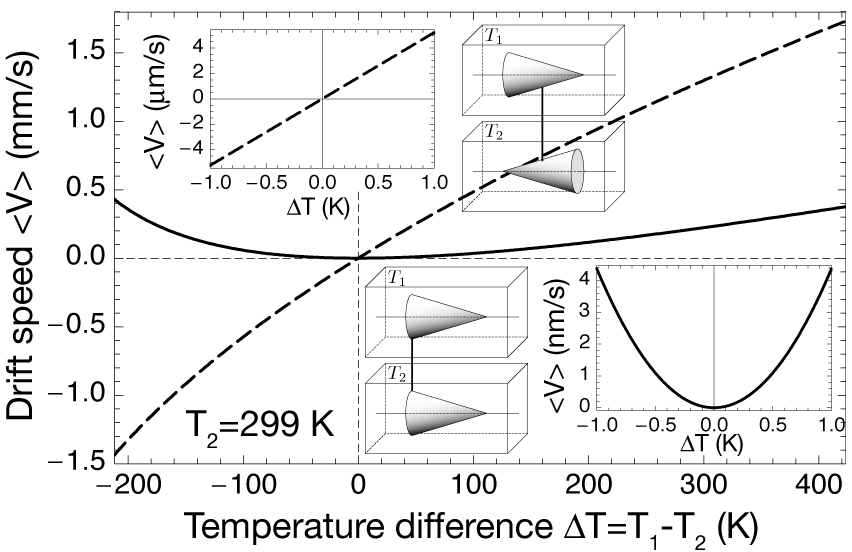}
\caption{The drift speed as a function of the temperature difference $\Delta T = T_{1} - T_{2}$ between the reservoirs for two configurations with identical motor units: parallel (solid curves and lower inset) and antiparallel (dashed curves and upper inset). The units are cone-shaped (opening angle 30$^{\circ}$) and the motor mass is 1000\,kDa. Both reservoirs are filled with argon gas of same density and $T_{2}$ is fixed at 299\,K. 
For small $\Delta T$ (see insets), the dependence is parabolic for the parallel setup and linear for the antiparallel setup. The drift speed is of order nm/s (parallel) and $\mu$m/s (antiparallel) for $\Delta T \approx 1\,$K.}
\label{fig:Vtemp}
\end{figure}
For simplicity we limit the discussion of the drift velocity, in particular in  relation to the applied temperature gradient, to the case of two reservoirs. Furthermore, the impact of the geometry is most clearly demonstrated when the motor units have an identical shape in both reservoirs.
There are then two possibilities, namely either the units have the same orientation (parallel), or they are pointing in opposite direction (antiparallel), see Fig.~\ref{fig:Vtemp} for a schematic representation.
For the first scenario, with $\sigma_{2,1} = \sigma_{2,2} = \sigma_{2}$ and $\sigma_{3,1} = \sigma_{3,2} = \sigma_{3}$, Eq.~(\ref{driftspeed}) yields:
\begin{equation}
\langle V \rangle_{e}
= \sqrt{\frac{m}{M}} \sqrt{\frac{\pi k_{B}}{8 M}}
\frac{\rho_{1} \rho_{2} (T_{1} - T_{2}) (\sqrt{T_{2}} - \sqrt{T_{1}})}{(\rho_{1} \sqrt{T_{1}} + \rho_{2} \sqrt{T_{2}})^{2}}
\frac{\sigma_{3}}{\sigma_{2}}.
\label{eq:scenario1}
\end{equation}
For the second scenario, careful consideration of the sign of $\cos\theta$ in Eq.~(\ref{eq:sigma}) leads us to write $\sigma_{2,1} = \sigma_{2,2} = \sigma_{2}$ as in the first scenario, but now when $\sigma_{3,1} = \sigma_{3}$, it follows that $\sigma_{3,2} = -\sigma_{3}$. The expression for the drift speed thus becomes:
\begin{equation}
\langle V \rangle_{o}
= \sqrt{\frac{m}{M}} \sqrt{\frac{\pi k_{B}}{8 M}}
\frac{\rho_{1} \rho_{2} (T_{1} - T_{2}) (\sqrt{T_{1}} + \sqrt{T_{2}})}{(\rho_{1} \sqrt{T_{1}} + \rho_{2} \sqrt{T_{2}})^{2}}
\frac{\sigma_{3}}{\sigma_{2}}.
\label{eq:scenario2}
\end{equation}
One striking feature of these results is that the drift velocity is scale-invariant. This is a general property: both $\sigma_{2}$ and $\sigma_{3}$ scale linearly with the total surface $S$ of the motor units.  As they appear in the denominator and the nominator respectively in Eqs.~(\ref{eq:scenario1}, \ref{eq:scenario2}), the scale dependence cancels out. This becomes even more apparent for the particular cases presented in Table \ref{tabel:sigma}, where $\sigma_{2} / S$ and $\sigma_{3} / S$ are expressed in topological terms. Note that the scale invariance of the drift speed is only valid with respect to the scale of the entire motor.  The relative proportions of separate motor units do matter.  Note also that scale invariance applies when disregarding the dependence on the mass $M$. For comparison with a physically  realistic situation, we assume a constant density of the motor, so that the drift velocity will decrease with increasing size of an object, through its $1/M$-dependence.

To investigate the departure from the equilibrium state, we consider small deviations of $T_{1}$ and $T_{2}$ about the average $T$,
\begin{equation}
T_{1} = T + \frac{\Delta T}{2}, \quad T_{2} = T - \frac{\Delta T}{2},
\end{equation}
so that for $\Delta T / T \ll  1$ the drift speeds tend to:
\begin{align}
\langle V \rangle_{e} &\rightarrow  
- \frac{1}{16} \sqrt{\frac{\pi}{2}}
\sqrt{\frac{m}{M}}
\frac{\rho_{1} \rho_{2}}
{(\rho_{1} + \rho_{2})^{2}}
\frac{\sigma_{3}}{\sigma_{2}}
\sqrt{\frac{k_{B} T}{M}}
\left(\frac{\Delta T}{T}\right)^{2},
\\
\langle V \rangle_{o} &\rightarrow  
\frac{1}{4} \sqrt{\frac{\pi}{2}}
\sqrt{\frac{m}{M}}
\frac{\rho_{1} \rho_{2}}
{(\rho_{1} + \rho_{2})^{2}}
\frac{\sigma_{3}}{\sigma_{2}}
\sqrt{\frac{k_{B} T}{M}}
\left(\frac{\Delta T}{T}\right).
\end{align}

We note that the respective orientation of the motor elements, parallel or antiparallel, 
 plays a crucial role. In the first case the sustained average displacement is always in the same direction (that of $- \sigma_{3}$), indifferent of the sign of $\Delta T$. This was to be expected because there is an additional symmetry in the system: the interchange of the temperature reservoirs has no effect. From the point of view of irreversible thermodynamics, this situation is special since there is no linear relation between the thermodynamic force (the temperature gradient) and flux (the resulting speed of the motor). For the second case of antiparallel alignment, one observes the usual
situation of linear response between thermodynamic force  and flux \cite{Reimann,vandenbroeckefficiency}:  
   equilibrium is a point of flux reversal, the direction of net motion reversing with $\Delta T$-inversion.

As an illustration, we reproduce, in table \ref{tabel:speedvalues}, explicit values for the drift speed in the case of a single
asymmetric unit, namely a cylindrically symmetric cone, positioned in antiparallel alignment in the two reservoirs under physically realistic conditions.
The degree of asymmetry is described by a single parameter, the opening angle $\alpha$. The geometric factors $\sigma_{2}$ and $\sigma_{3}$ are known analytically (see Table \ref{tabel:sigma}) and we find the following simple expression for the drift speed:
\begin{equation}
\langle V \rangle
= \sqrt{\frac{m}{M}} \sqrt{\frac{\pi k_{B}}{8 M}}
\frac{\rho_{1} \rho_{2} (T_{1} - T_{2}) (\sqrt{T_{1}} + \sqrt{T_{2}})}{(\rho_{1} \sqrt{T_{1}} + \rho_{2} \sqrt{T_{2}})^{2}}
(\sin \alpha - 1).
\label{eq:cone}
\end{equation}
The drift speed will become zero for $\alpha = 90^{\circ}$, namely when the cone loses its asymmetry and reduces to a flat disk.
A natural question is whether there is an optimal opening angle $\alpha_{o}$ that maximizes the drift speed. For a fixed cross section, one finds $\alpha_{o} = \sec^{-1}[\sqrt{(1 + \sqrt{5})/2}] \approx 38^{\circ}$. If, on the other hand, one assumes that the mass is kept constant,  a maximal speed is reached for an infinitely sharp cone.
\begin{table}
\caption{\label{tabel:speedvalues} Values for the drift speed of the motor, as predicted by theory. Realizations of the motor at different length scales from micrometers to nanometers are presented. The motor units are modeled as silica (SiO$_{2}$) cone-shaped objects, located in two reservoirs containing argon gas at temperatures 299.0 K and 299.1 K. For constant motor mass, the speed increases as the opening angle $\alpha$ of the cone decreases.}   
\begin{ruledtabular}  
\begin{tabular}{l  ll ll}
& $\alpha = 30^{\circ}$ & $\alpha = 5^{\circ}$ & $\alpha = 30^{\circ}$ & $\alpha = 5^{\circ}$ \\
Motor&  &  \\
mass& \multicolumn{2}{c}{Cylinder base} & \multicolumn{2}{c}{Drift speed} \\
(kDa) & \multicolumn{2}{c}{(nm)} & \multicolumn{2}{c}{($\mu$m/s)} \\
\hline
$10^{10}$ & 2600 & 1400 & 5.2$\times10^{-8}$ & 9.5$\times10^{-8}$\\
$10^{7}$ &  260 & 140 & 5.2$\times10^{-5}$ & 9.5$\times10^{-5}$\\
$10^{6}$ & 120 & 55 & 5.2$\times10^{-4}$ & 9.5$\times10^{-4}$\\
$10^{5}$ & 63 & 29 & 5.2$\times10^{-3}$ & 9.5$\times10^{-3}$\\
$10^{4}$ & 26 & 14 & 0.052 & 0.095\\
1000 & 12 & 6.3 & 0.52 & 0.95\\
100 & 5.5 & 2.9 & 5.2 & 9.5\\
10 & 2.6 & 1.4 & 52 & 95\\
\end{tabular}
\end{ruledtabular}
\end{table}
 
Note finally the very strong size-dependence: objects of 20 nm cover their length 5 times per second, for 5 nm size objects this becomes 1200 times per second. 

\section{Conclusion}

We have  calculated, on the basis of an exact microscopic theory, the properties of a thermal Brownian motor in a three-dimensional setup. When  detailed balance is broken by the application of a temperature gradient, a systematic net speed appears, as given in Eq. (\ref{driftspeed}). 
As an example, for a motor consisting of cone-shaped silica units of size 20 nm, one obtains a drift speed of about 0.1 $\mu$m/s when subject to   0.1 K temperature difference in a gaseous environment.
It remains to be seen whether the predictions of our theoretical analysis (involving various simplifications such as molecular chaos, elastic and normal interactions between gas particles and motor, expansion in mass ratio) provide a realistic estimate, especially for motors operating in a viscous environment.

\end{document}